# The Digital Pirahã Condition: Ecological Mismatch and the Reconstruction of Recursive Cognition

*Preprint version*

Dhushy Thillaivasan[a,*], Samar Shailendra[b], Kristina Nicholls[c], Deepani Guruge[d]

[a] *School of Business, Melbourne Institute of Technology, Melbourne, Australia*
[b] *School of IT & Engineering, Melbourne Institute of Technology, Melbourne, Australia*
[c] *School of Business, Melbourne Institute of Technology, Melbourne, Australia*
[d] *School of IT & Engineering, Melbourne Institute of Technology, Melbourne, Australia*

---

## Abstract

Contemporary digital and AI-mediated environments are reshaping the cognitive ecologies within which human reasoning develops. As everyday activity becomes embedded in datafied infrastructures, cognitive habits adapt to conditions of immediacy, fragmentation, externalisation, and algorithmic filtering. This paper introduces the Digital Pirahã Condition, a cultural-ecological model explaining how these environments cultivate adaptive but shallow cognitive patterns, epistemic flattening, reduced recursive capacity, and heightened reliance on external scaffolds. While functional within digital systems, these adaptations create an ecological mismatch with the recursive, integrative reasoning required in academic and institutional activity systems.

The paper argues that this mismatch is an ecological outcome rather than a psychological deficit, and that addressing it requires intentional cognitive niche construction within educational institutions. The lecturer is conceptualised as a cultural entrepreneur who reconstructs the cognitive ecology of learning through analog sanctuaries, AI-supported metacognitive scaffolds, and recursive curriculum architectures. The Digital Pirahã Condition thus provides a theoretical lens for understanding contemporary cognitive change and a framework for ecological redesign in AI-mediated societies.

* Corresponding author.

*Email addresses*: DThillaivasa@mit.edu.au (D. Thillaivasan), SShailendra@mit.edu.au (S. Shailendra), KNicholls@mit.edu.au (K. Nicholls), DGuruge@mit.edu.au (D. Guruge)

## 1. Introduction

Contemporary digital and AI-mediated environments increasingly shape the conditions under which human cognition develops. As everyday activity becomes embedded within *datafied* infrastructures, search engines, recommendation systems, predictive interfaces, and generative AI, cognitive processes unfold within environments that privilege speed, immediacy, fragmentation, and externalisation(Couldry & Mejias, 2019; Kitchin, 2017). These systems reorganise the sociotechnical ecology through which thinking is enacted(Gillespie, 2014). Research on digital reading and attention shows that high-velocity information flows and screen-based media encourage scanning, shallow processing, and reduced cognitive stamina (Baron, 2021, 2023; Wolf, 2018). Algorithmic feeds and pervasive cognitive offloading do not merely distract; they restructure the material and symbolic conditions that support reasoning(Beer, 2017; Zuboff, 2019). From a cognitive-ecological perspective, this matters because cognition emerges through participation in culturally and technologically organised activity (Clark, 2008; Hutchins, 1995). When the infrastructures that mediate activity change, the forms of reasoning that can develop shift accordingly.

This paper argues that widely observed difficulties in sustaining attention, integrating information across time, and engaging in recursive, hierarchical reasoning cannot be understood through individualistic or deficit-based explanations. Instead, they reflect a broader ecological transformation in how cognition is distributed, scaffolded, and enacted within *datafied and algorithmically mediated* environments. Research on cognitive offloading and distributed intelligence demonstrates that individuals increasingly rely on external systems to store, organise, and retrieve information (Clark & Chalmers, 1998; Risko & Gilbert, 2016). While such offloading is adaptive within many sociotechnical niches, it alters the developmental and cultural conditions under which recursive, integrative reasoning is cultivated (Aguera y Arcas, 2025; Davies, 2024).

To conceptualise this shift, the paper introduces the *Digital Pirahã Condition*, a cognitive-ecological model describing how digital environments cultivate cognitive habits optimised for immediacy, externalisation, and surface-level processing. These habits are adaptive within algorithmically accelerated ecologies but misaligned with cultural practices, such as deep reading, conceptual synthesis, and long-horizon decision-making, that depend on sustained recursion and the construction of hierarchical knowledge structures (Baron, 2021; Carr, 2011). The result is an ecological mismatch: cultural institutions continue to assume forms of cognition that the surrounding sociotechnical ecology no longer reliably supports (Laland & Brown, 2011; Sterelny, 2003).

The purpose of this paper is to diagnose this mismatch and outline how cultural institutions, particularly higher education institutions, can reconstruct the cognitive conditions necessary for recursive, integrative reasoning. The paper makes three contributions:

- It synthesises research in cognitive ecology, distributed intelligence, and digital cognition to explain how contemporary datafied environments reshape cognitive habits.
- It develops the Digital Pirahã Condition as a theoretical framework for understanding the collapse of recursion, epistemic flattening, and the rise of externally scaffolded cognition.
- It outlines an ecological redesign agenda for higher education, proposing institutional, curricular, and pedagogical interventions that rebuild the cognitive conditions required for deep learning.

The central research question guiding this paper is: How do datafied and AI-mediated environments reshape the cognitive ecology of learning, and what forms of ecological redesign are required for recursive, integrative reasoning to flourish?

To address this question, the paper proceeds as follows. Section 2 reviews foundational and contemporary research on cognitive ecology, distributed intelligence, and digital cognition. Section 3 develops the Digital Pirahã Condition as a theoretical model. Section 4 examines the societal and institutional implications of this ecological shift. Section 5 outlines an ecological redesign agenda for higher education, proposing interventions that rebuild the cognitive conditions necessary for recursive reasoning.

## 2. Background and Literature Review

The Digital Pirahã Condition is grounded in three intersecting bodies of scholarship: (1) Cultural evolution and cognitive niche construction, (2) Digital environments as high-velocity cognitive niches, and (3) Distributed intelligence and cognitive offloading.

Together, these literatures explain how contemporary digital ecologies reshape cognitive habits and why an ecological mismatch has emerged between the cognitive demands of cultural institutions and the cognitive habits cultivated by digital environments.

### 2.1 Cultural Evolution and Cognitive Niche Construction

Human cognition develops within culturally organised environments rather than emerging solely from biological evolution. Cultural evolution theory (Boyd & Richerson, 1985; Henrich, 2015) emphasises that humans are obligate cultural learners, dependent on socially transmitted knowledge to navigate complex environments. This dependence produces cognitive niche construction, in which cultural practices, tools, and institutions modify the selective pressures acting on cognition (Tomasello, 2014).

Henrich's (2015) and Muthukrishna's (2023) concept of the collective brain highlights how human success depends on the accumulation and transmission of knowledge across generations. Yet this same mechanism means that cognitive habits adapt to the structure of the surrounding cultural environment. When environments change rapidly, as in the digital era, cognitive strategies shift accordingly.

Cultural evolution, therefore, provides a foundation for understanding why individuals adapt to digital immediacy and why these adaptations may conflict with the recursive, temporally extended reasoning required in many cultural practices. These cultural-evolutionary dynamics provide the foundation for understanding why digital environments function as powerful cognitive niches that reshape reasoning, setting the stage for examining how high-velocity digital ecologies alter attentional and cognitive habits.

### 2.2 Digital Environments as High-Velocity Cognitive Niches

Digital information environments differ fundamentally from the slower, text-based ecologies that historically scaffolded deep, recursive cognition. Research on reading and attention (Baron, 2021, 2023; Wolf, 2018) shows that digital media encourage rapid scanning, shallow processing and reduced engagement with complex, linear argumentation. Cytowic (2024) similarly argues that constant sensory input and multitasking create attentional fragmentation, limiting the brain's capacity for sustained, recursive thought.

From a cognitive load perspective, high-velocity digital environments impose frequent task switching and interrupt working memory, reducing the ability to maintain hierarchical mental structures. These environments reward immediacy, novelty, and rapid retrieval rather than slow, integrative reasoning.

This shift does not reflect a decline in ability but an adaptive response to a niche where information is abundant, transient, and externally stored. Digital ecologies thus provide the environmental conditions that give rise to the Digital Pirahã Condition**.**

### 2.3 Distributed Intelligence, Cognitive Offloading, and the Arms Race of Recursion

The extended mind thesis (Clark & Chalmers, 1998) and research on cognitive offloading (Risko & Gilbert, 2016) demonstrate that humans routinely outsource memory, navigation, and computation to external systems. In digital environments, this offloading becomes pervasive: search engines store facts, algorithms filter information, and AI systems increasingly perform tasks that once required internal cognitive labour.

Davies (2024) argues that contemporary information systems generate structural overload that exceeds individuals' capacity for sustained modelling, pushing cognition toward rapid, surface-level responses rather than the recursive integration required for deeper understanding.

Agüera y Arcas (2022; 2025) contends that intelligence is becoming an arms race between organisms and their environments: as external systems grow more capable, humans must develop higher-order cognitive skills, abstraction, metacognition, and recursive self-modelling to remain effective participants in distributed intelligence systems.

Dennett's (2014) distinction between competence without comprehension provides an additional lens: digital environments increasingly enable task-level performance without the development of the recursive, self-monitoring structures that underpin genuine understanding.

The Digital Pirahã Condition can therefore be understood as a form of environmentally enabled competence that suppresses the emergence of higher-order recursive reasoning. Cultural institutions continue to assume that individuals will develop recursive capacities internally, while digital environments externalise cognition in ways that reduce opportunities to practise these skills.

### 2.4 Synthesis: The Ecological Foundations of the Digital Pirahã Condition

Across these literatures, a coherent picture emerges:

- Cultural evolution explains why cognition adapts to environmental structure.
- Digital environments constitute a new cognitive niche that rewards immediacy and fragmentation.
- Distributed intelligence systems shift cognitive labour outward, reducing opportunities for internal recursive development.

Together, these forces generate the ecological mismatch at the heart of the Digital Pirahã Condition: cultural practices continue to depend on recursive, integrative reasoning, while digital environments cultivate cognitive habits optimised for speed, externalisation, and surface-level processing.

While these ecological dynamics explain how cognitive habits adapt to digital environments, they do not yet account for the structural role of *datafied* platforms in actively configuring the cognitive conditions under which reasoning unfolds. Section 2.5 therefore, examines how algorithmic infrastructures shape contemporary cognitive ecologies.

### 2.5 Algorithmic Mediation and Datafied Cognitive Infrastructures

Contemporary digital environments are not only high-velocity cognitive niches; they are also *datafied sociotechnical infrastructures* that actively shape the conditions under which cognition unfolds (Couldry & Mejias, 2019; Smart et al., 2017). Research on datafication and algorithmic mediation shows that platforms do not simply transmit information but structure the flows, rhythms, and forms of engagement through which cognitive activity is organised (Gillespie, 2014; Kitchin, 2017). Recommender systems, ranking algorithms, and predictive interfaces curate attention, compress decision horizons, and privilege rapid, low-friction interactions over slow, deliberative reasoning (Zuboff, 2019). These infrastructures create deeply engineered cognitive ecologies in which immediacy and surface-level processing are not merely common but structurally incentivised, reshaping the ecological conditions for reflective autonomy (Beer, 2017; Seaver, 2019).

From a sociotechnical perspective, digital cognition is therefore produced within environments shaped by platform architectures, optimisation logics, and data extraction imperatives (Couldry & Mejias, 2019; Zuboff, 2019). The design of these systems, continuous scroll, push notifications, and hyper-personalised curation, creates attentional micro-ecologies that fragment cognitive continuity and reduce opportunities for recursive integration (Beer, 2017; Smart et al., 2017). Rather than acting as neutral tools, platforms function as active cognitive environments that systematically organise how information is encountered, how task-switching is forced, and how reflective reasoning unfolds over time (Seaver, 2019).

This perspective complements cognitive ecology by highlighting how the material and algorithmic infrastructures of digital platforms reshape the affordances available for recursive thought. When cognitive activity is mediated through systems optimised for engagement, prediction, and efficiency, the ecological conditions that support sustained modelling, abstraction, and hierarchical reasoning become increasingly rare. The Digital Pirahã Condition thus emerges not only from individual adaptations to digital immediacy but from the structural properties of *datafied* environments that systematically privilege shallow, externally scaffolded cognition.

## 3. The Digital Pirahã Condition: A Theoretical Framework

Section 2 established the ecological and sociotechnical foundations of contemporary cognition. Building on this, Section 3 formalises the Digital Pirahã Condition as a conceptual model that explains how digital and AI-mediated environments reshape cognitive activity. The term functions as a metaphorical ecological descriptor, not an anthropological claim about the Pirahã people. It captures a structural pattern: a shift toward immediacy, surface-level processing, and externally scaffolded cognition within digitally saturated environments.

The model comprises five interlocking components: (1) Immediate experience as a cognitive mode, (2) Collapse of recursion, (3) Epistemic flattening, (4) Ecological mismatch, and (5) Sanctuary logic as counter-ecological design.

Together, these components explain how digital ecologies cultivate cognitive habits optimised for speed and externalisation, and why these habits conflict with cultural practices that require slow, recursive, internally maintained reasoning.

### 3.1 Immediate Experience as a Cognitive Mode

Everett's (2005; 2009) account of the Pirahã describes a cultural orientation toward the Principle of Immediate Experience: privileging information that is perceptually available, socially witnessed, or directly relevant to the present moment. While debates continue regarding the linguistic implications (Nevins et al., 2009), the broader anthropological insight is widely accepted: cognitive preferences are shaped by cultural ecologies**.**

The Digital Pirahã Condition uses this as an analogy, not a linguistic claim. Digital platforms are structured around immediacy, novelty, rapid turnover, and algorithmic filtering. These environments reward cognitive habits optimised for speed rather than depth. Research on media multitasking shows that constant task switching reduces working memory capacity and impairs the formation of sustained, hierarchical representations (Ophir et al., 2009). Studies of digital reading demonstrate that screen-based environments encourage scanning and skimming rather than deep, linear engagement (Baron, 2021, 2023; Wolf, 2018). Work in platform studies similarly shows how algorithmic curation privileges immediacy and recency over depth (Beer, 2017; Seaver, 2019).

From a cognitive ecological perspective, digital environments constitute a new cognitive niche, one that scaffolds immediacy, externalisation, and rapid retrieval rather than the recursive, integrative reasoning required in many cultural practices. This shift is adaptive, not a deficit. The challenge arises when cultural practices require precisely the forms of cognition that digital ecologies suppress.

### 3.2 Collapse of Recursion

Recursive thinking, the ability to embed ideas within ideas, integrate multiple perspectives, and construct hierarchical mental models, is foundational to advanced reasoning. It underpins argumentation, theory building, conceptual synthesis, and long-horizon decision-making.

Cognitive scientists emphasise that recursion depends on sustained attention, internal representation, and slow, effortful processing (Hofstadter, 2008; Kahneman, 2012; Stanovich & West, 2000). Digital environments disrupt these conditions. They present information as notifications, short-form videos, search snippets, and algorithmically curated micro-content.

These formats encourage cognitive flattening, reducing complex ideas to surface-level cues. Frequent interruptions impose extraneous cognitive load, interrupting schema construction (Kirschner & van Merriënboer, 2013). Baron (2021) and Carr (2011) show that digital reading weakens the cognitive stamina required for recursion.

The collapse of recursion is therefore a structural ecological outcome, not a psychological failure. When environments rarely demand nested reasoning, recursive cognition atrophies.

### 3.3 Epistemic Flattening

Epistemic flattening refers to the shift from internalised knowledge structures to externally distributed fragments of information. In digital environments, individuals increasingly rely on external systems to store, retrieve, and organise knowledge. This aligns with research on cognitive offloading (Risko & Gilbert, 2016) and the extended mind thesis (Clark & Chalmers, 1998). Offloading is adaptive, but only when it is strategic. When it becomes habitual, individuals may access information without integrating it, retrieve facts without constructing coherence, and perform tasks without understanding the underlying structure.

Dennett's (2014) distinction between cranes and skyhooks clarifies this dynamic. External supports can function as scaffolds that build cognitive structure, or as shortcuts that bypass the recursive processes required for genuine understanding. Digital environments increasingly operate as skyhooks**:** enabling performance while undermining the development of internal coherence. This produces a paradox: students feel knowledgeable precisely because the environment allows them to avoid the cognitive labour required to become knowledgeable (Baron, 2021; Bjork et al., 2013).

The Digital Pirahã Condition captures this shift: a move from recursive, internally maintained representations toward flattened, externally distributed fragments.

### 3.4 Ecological Mismatch

Cultural and higher education institutions, research communities, and professional fields rest on implicit assumptions about the cognitive ecology of their participants, sustained attention, recursive reasoning and internalised conceptual frameworks. These assumptions reflect a pre-digital cognitive niche in which information was scarce, slow to access, and costly to retrieve. Under such conditions, internalisation and synthesis were adaptive strategies (Sterelny, 2003; Tomasello, 2014).

Digital environments invert these pressures. Information is abundant, instantaneous, and externally stored. The adaptive strategy becomes rapid scanning, selective attention and reliance on distributed systems. This creates an ecological mismatch between the cognitive habits shaped by digital culture and the recursive reasoning required in many cultural practices. Evolutionary mismatch theory explains how previously adaptive traits can become maladaptive when environments change rapidly (Laland & Brown, 2011).

Sanctuary logic emerges as a response to this mismatch: the intentional construction of counter-ecologies that restore the conditions necessary for recursive reasoning.

### 3.5 Sanctuary Logic as Counter-Ecological Design

If digital environments constitute a cognitive niche that rewards immediacy, fragmentation, and externalisation, then cultural institutions must intentionally construct counter-niches that support recursive, internally coherent reasoning. This is the logic of sanctuaries: ecological interventions that temporarily suspend the pressures of digital environments to enable forms of cognition that cannot flourish under conditions of constant interruption.

### 3.5.1 Cognitive (Analog) Sanctuaries

Device-free environments that support sustained attention, deep reading and hierarchical reasoning are supported by research on attention, working memory, and reading comprehension (Baron, 2021, 2023; Wolf, 2018; Ward et al., 2017).

### 3.5.2 AI Sanctuaries

AI is used not as a substitute for thinking, but as a metacognitive scaffold**,** generating counterarguments, prompting reflection, supporting conceptual mapping, and revealing hidden assumptions, aligned with emerging research on AI-supported learning (Dalsgaard, 2025; Holmes & Tuomi, 2022)

### 3.5.3 Sanctuaries as Counter-Ecologies

Sanctuaries are not nostalgic retreats. They are designed cognitive ecologies that reconstruct the conditions necessary for recursive reasoning. They complete the Digital Pirahã model by showing how institutions can counteract epistemic flattening and ecological mismatch through intentional ecological redesign.

Figure #1 depicts the Digital Pirahã Model as a Cognitive Ecology. The figure illustrates the progression from digital cognitive ecology through cognitive adaptation and academic demands to ecological counter-design in higher education, culminating in restored recursive, integrative, and metacognitive reasoning.

**Figure #1:**
*The Digital Pirahã Model as a Cognitive Ecology*

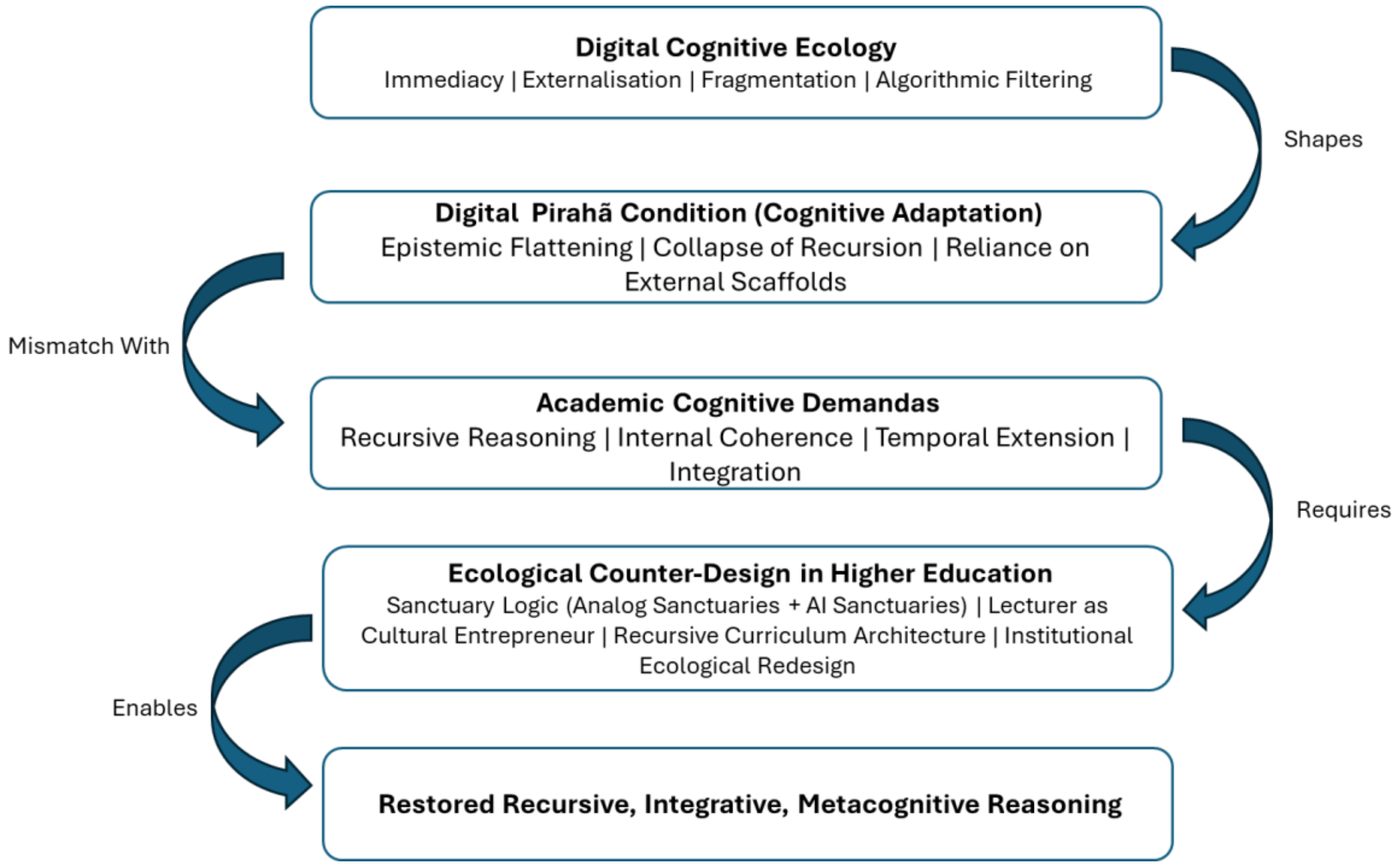

*Note.* This figure depicts the Digital Pirahã Model as a cultural-ecological account of contemporary cognition. AI-mediated and digitally saturated environments constitute a

cognitive niche characterised by immediacy, externalisation, fragmentation, and algorithmic filtering. These conditions cultivate adaptive cognitive habits, epistemic flattening, reduced recursive capacity, and increased reliance on external scaffolds that together form the Digital Pirahã Condition. This condition stands in structural tension with cultural practices that depend on recursive, integrative, and temporally extended reasoning. The model also identifies a set of ecological counter-designs, analog sanctuaries, AI-supported metacognitive scaffolds, the lecturer as cultural entrepreneur, recursive curriculum architecture, and institutional redesign, that reconstruct the cognitive conditions necessary for restored recursive, integrative, and metacognitive reasoning.

## 4. Societal Implications

The Digital Pirahã Condition reveals a structural mismatch between the cognitive habits cultivated by digital environments and the recursive, integrative reasoning required in academic and professional activity systems. Addressing this mismatch requires more than pedagogical technique; it demands intentional cognitive niche construction within educational institutions. This section conceptualises the lecturer as a cultural entrepreneur who reconstructs the cognitive ecology of learning, reintroducing the friction, depth, and temporal extension that digital environments suppress.

This ecological framing aligns with research on distributed cognition (Hutchins, 1995), niche construction (Sterelny, 2003), and the extended mind (Clark & Chalmers, 1998), all of which emphasise that cognitive capacities develop within structured environments. In the context of the Digital Pirahã Condition, the lecturer becomes a designer of counter-ecologies that resist epistemic flattening and cultivate recursive reasoning.

### 4.1 The Lecturer as a Cultural Entrepreneur

Mokyr's (2017) concept of the cultural entrepreneur describes actors who reshape the cultural menu available to a community. Within higher education, lecturers occupy a similar role: they curate intellectual environments, model cognitive practices, and shape the norms of inquiry that govern participation in academic activity systems.

When digital environments select for immediacy, fragmentation, and cognitive offloading (Risko & Gilbert, 2016), the lecturer becomes one of the few remaining agents capable of sustaining a culture of slow, recursive reasoning. This role is not about charisma but about environmental design: constructing learning spaces that reward depth over speed and coherence over retrieval.

In this sense, the lecturer mediates between two ecologies: the digital niche that rewards immediacy, and the academic niche that requires recursion.

### 4.2 Reintroducing Cognitive Friction

A central function of the cultural entrepreneur is to reintroduce productive cognitive friction. Taleb's (2012) concept of antifragility suggests that systems grow stronger when exposed to manageable stressors. In educational contexts, such stressors include complex texts, extended arguments and tasks requiring integration across time.

Digital environments minimise friction through instant retrieval, algorithmic simplification, and constant interruption (Carr, 2011; Ophir et al., 2009). Academic reasoning, by contrast, requires learners to tolerate ambiguity, navigate complexity, and construct internal coherence.

By designing tasks that demand recursive reasoning, multi-stage argumentation, comparative analysis, and theoretical synthesis, the lecturer restores the cognitive demands that digital environments erode. This friction is developmental, not punitive: it rebuilds the cognitive stamina required for deep learning.

### 4.3 Designing Cognitive (Analog) Sanctuaries

One practical expression of cognitive niche construction is the creation of cognitive (analog) sanctuaries: structured periods or spaces in which digital devices are intentionally excluded to support sustained, uninterrupted thought.

Research shows that even the mere presence of a smartphone can impair cognitive performance (Ward et al., 2017). Analog sanctuaries remove this background load, enabling learners to engage in the slow, recursive processing that digital environments disrupt (Baron, 2021, 2023; Wolf, 2018). Examples include device-free seminars, handwritten synthesis tasks, slow reading sessions and silent reflection periods.

These sanctuaries operationalise the sanctuary logic introduced in Section 3.5, functioning as counter-ecologies that cultivate the cognitive capacities digital environments suppress.

### 4.4 Integrating AI as a Cognitive Scaffold Rather Than a Substitute

While analog spaces are essential, contemporary academic activity systems also involve ubiquitous AI tools. The challenge is to integrate AI not as a substitute for thinking but as a metacognitive scaffold. AI can support recursive reasoning by generating alternative explanations for evaluation, providing counterarguments, supporting conceptual mapping and outlining and prompting reflective interrogation.

This aligns with the extended mind thesis (Clark & Chalmers, 1998) and emerging research on AI-supported learning (Holmes & Tuomi, 2022). In this framing, AI becomes a cognitive exoskeleton, enabling learners to externalise and refine their thinking without outsourcing the core cognitive labour.

This approach also aligns with the arms race model of intelligence (Aguera y Arcas, 2025): as AI automates lower-level tasks, humans must specialise in metacognition, judgement, and integration.

### 4.5 The Lecturer’s Range and the Problem of Cognitive Asymmetry

A further dimension of ecological reconstruction concerns the lecturer’s Range, their breadth of knowledge, cross-domain synthesis, and capacity to draw connections across disparate ideas. Epstein (2021) argues that individuals with broad integrative capacity are better equipped to navigate complex environments. In higher education, Range functions as a live demonstration of recursive cognition.

However, Range is unevenly distributed across the academic workforce, shaped by disciplinary silos, publication pressures, and institutional incentives that reward depth over breadth. This produces cognitive asymmetry between lecturers with integrative capacity and students whose cognitive habits have been shaped by epistemic flattening.

Students accustomed to fragmented digital information may react with surprise or intimidation when encountering a lecturer who demonstrates wide-ranging synthesis. This is not a failure of teaching but a predictable outcome of the Digital Pirahã ecology. To address this asymmetry, lecturers must make their cognitive processes transparent by narrating how connections are formed, explaining why analogies are chosen, and revealing steps of conceptual integration and modelling uncertainty and revision

This transparency transforms Range from an intimidating display into an accessible scaffold, enabling students to internalise the recursive habits they observe.

## 5. Conclusion

This paper set out to explain how AI-mediated and digitally saturated environments reshape contemporary cognition, and why these shifts generate a structural mismatch with the recursive, integrative reasoning required in academic and cultural activity systems. Drawing on research in cognitive ecology, distributed intelligence, and ecological mismatch, the paper developed the Digital Pirahã Condition as a cultural-ecological model of adaptive cognitive change.

The analysis demonstrates that digital environments constitute a high-velocity cognitive niche characterised by immediacy, fragmentation, externalisation, and algorithmic filtering. These ecological pressures cultivate adaptive cognitive habits, epistemic flattening, reduced recursive capacity, and increased reliance on external scaffolds that are functional within digital systems but misaligned with the demands of academic and institutional reasoning. This mismatch is not a psychological deficit but an ecological outcome: cognition adapts to the environments that shape it.

The paper further argued that addressing this mismatch requires intentional cognitive niche construction within educational and epistemic institutions. The lecturer emerges as a cultural entrepreneur who reconstructs the cognitive ecology of learning through analog sanctuaries, AI-supported metacognitive scaffolds, and recursive curriculum architectures. These counter-ecologies restore the conditions necessary for sustained, integrative, and temporally extended reasoning.

The societal implications are significant. As digital environments increasingly mediate attention, memory, and judgement, the capacity for recursive reasoning becomes a form of cultural resilience. Institutions that fail to recognise the ecological nature of cognition risk reinforcing epistemic flattening and undermining the foundations of knowledge production, democratic deliberation, and organisational decision-making. Conversely, institutions that design for recursion can cultivate the cognitive capacities required to navigate AI-mediated futures.

Future research should examine how different digital ecologies shape cognitive development across age groups, disciplines, and cultural contexts; how AI systems can be designed to support rather than substitute recursive reasoning; and how institutional architectures can be re-engineered to sustain deep cognitive practices in an era of pervasive externalisation.

The Digital Pirahã Condition thus offers a theoretical lens for understanding contemporary cognitive change and a practical framework for ecological redesign. By recognising cognition as environmentally situated and culturally scaffolded, we can better design the counter-ecologies necessary to sustain recursive, integrative, and metacognitive reasoning in a digital age.

## 6. Reference

Agüera y Arcas, B. (2022). Do Large Language Models Understand Us? *Daedalus*, *151*(2), 183–197. https://doi.org/10.1162/daed_a_01909

Aguera y Arcas, Blaise. (2025). *What is intelligence? : lessons from AI about evolution, computing, and minds*. The MIT Press ; Berggruen Institute.

Baron, N. S. . (2021). *How we read now : Strategic choices for print, screen, and audio*. Oxford University Press.

Baron, N. S. . (2023). *Who wrote this? : how AI and the lure of efficiency threaten human writing*. Stanford University Press.

Beer, D. (2017). The social power of algorithms. *Information Communication and Society*, *20*(1), 1–13. https://doi.org/10.1080/1369118X.2016.1216147

Bjork, R. A., Dunlosky, J., & Kornell, N. (2013). Self-Regulated Learning: Beliefs, Techniques, and Illusions. *Annual Review of Psychology*, *64*(1), 417–444. https://doi.org/10.1146/annurev-psych-113011-143823

Boyd, Robert., & Richerson, P. J. . (1985). *Culture and the evolutionary process*. University of Chicago Press.

Carr, N. G. . (2011). *The shallows : What the Internet is doing to our brains*. W.W. Norton.

Clark, A. (2008). *Supersizing the Mind: Embodiment, Action, and Cognitive Extension*. Oxford University PressNew York. https://doi.org/10.1093/acprof:oso/9780195333213.001.0001

Clark, A., & Chalmers, D. (1998). The Extended Mind. *Analysis*, *58*(1), 7–19. https://doi.org/10.1093/analys/58.1.7

Couldry, Nick., & Mejias, U. Ali. (2019). *The costs of connection : how data is colonizing human life and appropriating it for capitalism*. 323.

Cytowic, R. E. . (2024). *Your Stone Age brain in the screen age : Coping with digital distraction and sensory overload*. The MIT Press.

Davies, Dan. (2024). *The Unaccountability Machine: Why Big Systems Make Terrible Decisions—and How the World Lost Its Mind*. Profile Books Ltd.

Dennett, D. C. . (2014). *Intuition pumps and other tools for thinking*. Penguin Books.

Epstein, D. J. . (2021). *Range : Why generalists triumph in a specialized world*. Riverhead Books.

Everett, D. L. (2005). Cultural Constraints on Grammar and Cognition in Pirahã. *Current Anthropology*, *46*(4), 621–646. https://doi.org/10.1086/431525

Everett, D. Leonard. (2009). *Don't sleep, there are snakes : Life and language in the Amazonian jungle*. Vintage Departures.

Gillespie, T. (2014). The Relevance of Algorithms. In *Media Technologies* (pp. 167–194). The MIT Press. https://doi.org/10.7551/mitpress/9042.003.0013

Henrich, J. (2015). *The Secret of Our Success: How Culture Is Driving Human Evolution, Domesticating Our Species, and Making Us Smarter*. Princeton University Press. https://www.goodreads.com/book/show/25761655-the-secret-of-our-success

Hofstadter, D. R. . (2008). *I am a strange loop*. Basic Books.

Holmes, W., & Tuomi, I. (2022). State of the art and practice in education. *European Journal of Education*, *57*(4), 542–570. https://doi.org/10.1111/ejed.12533

Hutchins, Edwin. (1995). *Cognition in the wild*. MIT Press.

Kahneman, D. (2012). *Thinking,Fast and Slow*. Penguin Press.

Kirschner, P. A., & van Merriënboer, J. J. G. (2013). Do Learners Really Know Best? Urban Legends in Education. *Educational Psychologist*, *48*(3), 169–183. https://doi.org/10.1080/00461520.2013.804395

Kitchin, R. (2017). Thinking critically about and researching algorithms. *Information, Communication & Society*, *20*(1), 14–29. https://doi.org/10.1080/1369118X.2016.1154087

Laland, K. N., & Brown, G. (2011). *Sense and Nonsense: Evolutionary Perspectives on Human Behaviour*. Oxford University Press, USA.

Mokyr, J. (2017). A Culture of Growth: The Origins of the Modern Economy. In *A Culture of Growth*. Princeton University Press.

Muthukrishna, M. (2023). *A Theory of Everyone: The New Science of Who We Are, How We Got Here, and Where We're Going*. The MIT Press.

Nevins, A., Pesetsky, D., & Rodrigues, C. (2009). Pirahã exceptionality: A reassessment. *Language*, *85*(2), 355–404. https://doi.org/10.1353/lan.0.0107

Ophir, E., Nass, C., & Wagner, A. D. (2009). Cognitive control in media multitaskers. *Proceedings of the National Academy of Sciences*, *106*(37), 15583–15587. https://doi.org/10.1073/pnas.0903620106

Risko, E. F., & Gilbert, S. J. (2016). Cognitive Offloading. *Trends in Cognitive Sciences*, *20*(9), 676–688. https://doi.org/10.1016/j.tics.2016.07.002

Seaver, N. (2019). Captivating algorithms: Recommender systems as traps. *Journal of Material Culture*, *24*(4), 421–436. https://doi.org/10.1177/1359183518820366;REQUESTEDJOURNAL:JOURNAL:MCUA;WEBSITE:WEBSITE:SAGE;WGROUP:STRING:PUBLICATION

Smart, P., Heersmink, R., & Clowes, R. W. (2017). The cognitive ecology of the internet. *Cognition beyond the Brain: Computation, Interactivity and Human Artifice, Second Edition*, 251–282. https://doi.org/10.1007/978-3-319-49115-8_13

Stanovich, K. E., & West, R. F. (2000). Individual differences in reasoning: Implications for the rationality debate? *Behavioral and Brain Sciences*, *23*(5), 645–665. https://doi.org/10.1017/S0140525X00003435

Sterelny, Kim. (2003). *Thought in a hostile world : the evolution of human cognition*. Blackwell Pub.

Taleb, N. Nicholas. (2012). *Antifragile : Things that gain from disorder*. Random House.

Tomasello, Michael. (2014). *A natural history of human thinking*. Harvard University Press.

Ward, A. F., Duke, K., Gneezy, A., & Bos, M. W. (2017). Brain Drain: The Mere Presence of One's Own Smartphone Reduces Available Cognitive Capacity. *Journal of the Association for Consumer Research*, *2*(2), 140–154. https://doi.org/10.1086/691462

Wolf, Maryanne. (2018). *Reader, come home : The reading brain in a digital world*. Harper, an imprint of HarperCollinsPublishers.

Zuboff, Shoshana. (2019). *The age of surveillance capitalism : The fight for a human future at the new frontier of power*. PublicAffairs.